\documentclass[12pt]{article}
\usepackage{setspace}
\usepackage{amsmath,amssymb,mathrsfs}
\usepackage{color}
\usepackage{hyperref}
\numberwithin{equation}{section}
\begin{document}
\setlength{\topmargin}{-1cm} 
\setlength{\oddsidemargin}{-0.25cm}
\setlength{\evensidemargin}{0cm}
\newcommand{\e}{\epsilon}
\newcommand{\beq}{\begin{equation}}
\newcommand{\eeq}[1]{\label{#1}\end{equation}}
\newcommand{\bea}{\begin{eqnarray}}
\newcommand{\eea}[1]{\label{#1}\end{eqnarray}}
\renewcommand{\Im}{{\rm Im}\,}
\renewcommand{\Re}{{\rm Re}\,}
\newcommand{\diag}{{\rm diag} \, }
\newcommand{\Tr}{{\rm Tr}\,}
\newcommand{\rj}[1]{\textcolor{green}{#1}}
\def\draftnote#1{{\color{red} #1}}
\def\bldraft#1{{\color{blue} #1}}
\setcounter{tocdepth}{2}\begin{titlepage}
\begin{center}

\vskip 4 cm

{\Large \bf Lorentz Covariant Supertranslation Frames for the Angular Momentum Aspect}

\vskip 1 cm

{Reza Javadinezhad$^a$ \footnote{E-mail: \href{javadinezhad@ipm.ir}{Javadinezhad@ipm.ir} } and Massimo Porrati$^b$ \footnote{E-mail: \href{mailto:mp9@nyu.edu}{mp9@nyu.edu}} }

\vskip .75 cm

{\em $a)$ School of Physics, Institute for Research in Fundamental Sciences (IPM), \\P.O.Box 19395-5531, Tehran, Iran}
\vskip .75 cm

{\em $b)$ Center for Cosmology and Particle Physics, \\ Department of Physics, New York University, \\ 726 Broadway, New York, NY 10003, USA}

\end{center}

\vskip 1.25 cm

\begin{abstract}
\noindent  
In this letter, we review the well known ambiguity in defining angular momentum (and mass dipole) fluxes in general relativity and 
we reinterpret recent works that resolve the ambiguity by defining invariant charges. We resolve the ambiguity by finding the
conditions that fix a frame for supertranslation and for space-time translation. We present an elementary method for 
measuring the angular momentum aspect. We clarify the difference between
supertranslation invariant Lorentz charges and the generators of BMS
coordinate
transformations. We also work out explicitly the supertranslation frame-fixing conditions for the metric created by point particles to first nontrivial order in the Newton constant. 
\end{abstract}
\end{titlepage}
\newpage
\section{Introduction and review of Lorentz charges, aspects and fluxes}

The problem of defining angular momentum in general relativity is due to an abundance of choices. The  symmetry
 of asymptotically flat space contains the BMS (Bondi-van der Burg-Sachs-Metzner) algebra~\cite{bond1,bms,s1,s2}. This algebra
  contains an infinite dimensional Abelian subalgebra (the supertranslations) that extends the Poincar\'e algebra. The generators of
 Lorentz transformations $J_{\mu\nu}$ do not commute with supertranslations, so we end up with an infinite number of equally 
 good forms for angular momentum and boost (also called mass dipole). The flux of angular momentum through future null infinity
 $\mathscr{I}^+$ too suffers from this ambiguity. If we interpret the flux of angular momentum as due to an outgoing flux of physical
 (massless) particles, the ambiguity seems strange because all the different fluxes correspond to the same energy flux.
 This implies that angular momentum can be carried away by zero energy particles. This is not a contradiction but it 
 underscores the need of a proper 
 definition of angular momentum, mass dipole, and fluxes thereof~\cite{pen}. This can be done in two ways,  either by constructing
 invariant charges or by fixing a supertranslation frame. To preserve Lorentz invariance the prescription should be covariant. 
 This is the aim of this letter, which begins with setting notations and reviewing the construction of various charges and
 fluxes. It continues in section 2, which shows how the angular momentum aspect is a physical, measurable quantity subject to an
 equally physical ambiguity related to the choice of the supertranslation frame. The ambiguity is described in section 3. Section 4 gives a 
 complete, physical, well-defined prescription to fix the ambiguity for Lorentz charges, while preserving their covariance. This is the
  new part of~\cite{3puzz}, which we interpret in a different but equivalent way here as a frame-fixing procedure.
  Section 5 extends the prescription to the angular momentum and mass dipole {\em fluxes}; it also discusses how different, equally acceptable prescriptions --which corresponds to different measurement prescriptions-- change the value of the measured  fluxes and can be used to remove an infrared divergence in the mass dipole flux arising at $O(G^4)$ in the Newton constant $G$.
  Section 6 discusses the difference between the invariant Lorentz charges that
  have appeared in the literature and the generators of Lorentz
  transformations induced by asymptotic symmetries of the metric.
  Section 7 illustrates the Lorentz-covariance transformation properties of supertranslation frames by performing an explicit computation in the case of the metric created to  $O(G)$ by $n$ massive point particles. 

The metric near future null infinity $\mathscr{I}^+$ in Bondi-Sachs coordinates is~\cite{bond1,bms,s1,s2}
\bea
 ds^2  &=&  -du^2 - 2 du\, dr + r^2 \left(h_{AB}
+\frac{C_{AB}}{r}\right) d\Theta^A d\Theta^B + D^AC_{AB}\, du\, d\Theta^B +
\frac{2M}{r} du^2  \nonumber \\ && 
		+ \frac{1}{16 r^2} C_{AB}C^{AB} du dr 
	 + \frac{1}{r}\left(
		\frac{4}{3}\left(N_A+u \partial_A M \right) - \frac{1}{8} \partial_A \left(C_{BD}C^{BD}\right)
		\right) du d\Theta^A \nonumber \\ &&
		+ \frac{1}{4} h_{AB} C_{CD}C^{CD}d \Theta^A d \Theta^B
		+ \dots , 
		\eea{1}  
where the mass aspect $M(\Theta,u)$ is a scalar, the angular momentum aspect $N_A(\Theta,u)$ is a vector and the shear $C_{AB}(\Theta,u) $ is a symmetric and traceless tensor. All these quantities are defined on 
the celestial sphere with coordinates $\Theta_A$ and round metric $h_{AB}$ and also depend on the retarded time $u$. 
The dots in~\eqref{1} denote subdominant terms in $1/r$. 
The coordinate system in eq.~\eqref{1} is invariant under an asymptotic symmetry acting on the retarded time as $u\rightarrow u'=u+ f(\Theta)$, called supertranslation~\cite{bms}. It acts on the shear as
\beq
C_{AB}(\Theta,u)\rightarrow C'_{AB}(\Theta,u') = C_{AB}(\Theta,u) + 2 D_AD_B f -  h_{AB} D^2 f .
\eeq{1a}
It also acts on the angular coordinates as
$\Theta^A\rightarrow \Theta'^A= \Theta^A - r^{-1} D^A f(\Theta)$. 
In nonradiative regions where the Bondi news $N_{AB}=\partial_u C_{AB}$
vanishes, mass 
and  angular momentum aspect transform as
\bea
&& N_A(\Theta,u) \rightarrow N'_A(\Theta,u')=
N_A(\Theta,u)- 3M(\Theta,u)D_A f(\Theta) - D_AM(\Theta,u)  f(\Theta)  ,
\nonumber \\ 
&& M(\Theta,u)\rightarrow M'(\Theta,u')=M(\Theta,u) .
\eea{m1}
Here we set $\Theta=\Theta'$ since we are interested in the transformations
at leading order in $1/r$.

The transformation in radiative regions where the Bondi news tensor is nonzero
is much more complicated and we
refer to~\cite{yau-trans} for its explicit form.
In nonradiative regions $M$ and $N_A$ are independent
of $u$, so eq.~\eqref{m1} further simplifies to
$N'_A(\Theta,u+f(\Theta))= N_A(\Theta)- 3M(\Theta)D_Af(\Theta) -D_AM(\Theta) f(\Theta)$. 

Energy and Lorentz charges 
are defined in terms of $M$, $N_A$ and the six conformal Killing vectors $Y^A$ of the celestial sphere~\cite{bms}
\beq
E(u)={1\over 4\pi G} \int d^2 \Theta \sqrt{h} M(\Theta,u), \quad J_Y(u) = {1\over 8\pi G} \int d^2 \Theta \sqrt{h} Y^A N_A(\Theta,u).
\eeq{2}
By definition the conformal Killing vectors obey $D_A Y_B + D_B Y_A= h_{AB} D_CY^C $. For rotations $D_CY^C=0$ while for boosts
$D_CY^C$ is an $l=1$ harmonic. In the latter case we can write $Y_A=D_A\psi$ with $\psi$ obeying $D_AD^A \psi = -2 \psi$. 
Transformation laws~\eqref{m1} show that the  total energy flux $\Delta E \equiv E(+\infty)-E(-\infty)$ is invariant under supertranslations. This is not true 
for the flux of the Lorentz charges. In particular, the angular momentum flux
$\Delta J_Y \equiv J_Y(+\infty)-J_Y(-\infty)$ can be changed by a supertranslation, as noticed already in~\cite{pen}. 

Supertranslation-invariant definitions of the angular momentum flux for finitely radiating systems were given in~\cite{comp} and later, 
independently in~\cite{yau21a,yau21b}. Ref.~\cite{yau21b} also defines supertranslation-invariant Lorentz boost charges and their flux.
A different approach to charges and fluxes appears in~\cite{jkp}, where the flux of  invariant Lorentz charges is defined in terms of bulk 
integrals of soft-graviton dressed canonical degrees of freedom. The definition of the angular momentum flux given in~\cite{jkp} was later supplemented in~\cite{jp} with an additional supertranslation-invariant term that allowed to reproduce the flux computed in 
gravitational scattering to order $G^2$ ($G=$Newton's constant). 
Finally,~\cite{3puzz} supplements the definitions of~\cite{yau21a,yau21b} with the first example of fully Lorentz-covariant 
prescription for some redundant degrees of freedom introduced there.

In this letter, we describe a simpler albeit less elegant definition of the supertranslation-invariant Lorentz charges and fluxes of~\cite{yau21a,yau21b,3puzz}, which we hope may clarify their physical meaning.  
The link between fluxes integrated over null infinity $\mathscr{I}^+$ and the integrals on the future and past ends of null infinity, $\mathscr{I}^+_\pm$, given in~\eqref{2}  is obtained by using the constraint equations 
\bea
\partial_u M &=& -\frac{1}{8} N_{AB}N^{AB}+\frac{1}{4}D_AD_B N^{AB},\nonumber\\
\partial_u N_A &=& -\frac{1}{4} D^B \left( D_BD^CC_{CA}-D_AD^CC_{CB}\right)  -u \partial_u \partial_A M -T_{uA}, 
\nonumber  \\
T_{uA} &=& -\frac{1}{4}\partial_A \left(C_{BD}N^{BD}\right) + 
\frac{1}{4} D_B \left(C^{BC}N_{CA}\right) -\frac{1}{2}C_{AB}D_C N^{BC} , \quad N_{AB}\equiv \partial_u C_{AB} .
\eea{m2}

We could use~\eqref{m2} to find the canonical generators of Lorentz transformations, as it was done e.g. in~\cite{jkp}, but we
won't repeat the construction here. We just point out that conservation of angular momentum is just one of infinitely many
conservations laws implied by the matching conditions of the angular momentum aspect. 
To define them we need the asymptotic metric near past null infinity $\mathscr{I}^-$ 
\bea
 ds^2  &=&  -dv^2 + 2 dv\, dr + r^2 \left(h_{AB}
+\frac{\tilde{C}_{AB}}{r}\right) d\theta^A d\theta^B + D^A\tilde{C}_{AB}\, du\, d\theta^B +
\frac{2\tilde{M}}{r} dv^2  \nonumber \\ && 
		+ \frac{1}{16 r^2} \tilde{C}_{AB}\tilde{C}^{AB} du dr 
	 + \frac{1}{r}\left(
		\frac{4}{3}\left(\tilde{N}_A+v \partial_A \tilde{m} \right) - \frac{1}{8} \partial_A \left(\tilde{C}_{BD}\tilde{C}^{BD}\right)
		\right) du d\theta^A \nonumber \\ &&
		+ \frac{1}{4} h_{AB} \tilde{C}_{CD}\tilde{C}^{CD}d \theta^A d \theta^B
		+ \dots , 
\eea{m3}

The matching conditions, which imply conservation of charges, are then 
$\lim_{v\rightarrow +\infty} \tilde{N}_A(\theta,v)= \lim_{u\rightarrow -\infty} N_A(\Theta,u)$,
where $\theta$ and $\Theta$ are matched antipodally~\cite{klps}.

\section{Physical meaning of $N_A$}
Eqs.~\eqref{2} show that the gradient part of $N_A$ does not contribute to the angular momentum, because the Hodge decomposition
$N_A=D_A \phi + \epsilon_{AB}D^B \psi$ and the property $D_A Y^A=0$, which holds for rotation Killing vectors imply 
\bea
J_Y(u)  &=&
 {1\over 8\pi G} \int d^2 \Theta \sqrt{h} Y^A (D_A \phi + \epsilon_{AB}D^B \psi )=
- {1\over 8\pi G} \int d^2 \Theta \sqrt{h} (\phi D_A Y^A   + \psi \epsilon_{AB}D^B Y^A)  \nonumber \\
&=& - {1\over 8\pi G} \int d^2 \Theta \sqrt{h} \psi \epsilon_{AB}D^B Y^A .
\eea{m3aaa}

The curvature $D_A N_B - D_B N_A$ in nonradiative regions is measured by the relative time delay
 of two light rays moving in opposite directions along a closed path in the $\Theta$ coordinates.  At fixed $r$
a light ray moving in the metric~\eqref{1} obeys the equation
\beq
-\left({du \over d\lambda}\right)^2 + 2P_A  {d u \over d\lambda}{ d\Theta^A \over d\lambda} + Q_{AB} { d\Theta^A \over d\lambda}{ d\Theta^B \over d\lambda}=0
\eeq{m4}
where $\lambda$ is the affine parameter and 
\bea
2P_A &\equiv& \left( 1- {2M\over r}\right)^{-1} \left[ D^BC_{AB} +\frac{1}{r}\left(\frac{4}{3}\left(N_A+u \partial_A M \right) - \frac{1}{8} \partial_A \left(C_{BD}C^{BD}\right) \right) \right] \nonumber \\ &=& 
D^BC_{AB} + \frac{1}{r}\left(\frac{4}{3}\left(N_A+u \partial_A M \right) 
+ 2M D^BC_{AB} - \frac{1}{8} \partial_A \left(C_{BD}C^{BD}\right) \right) + O(r^{-2}), 
\nonumber \\
Q_{AB} &\equiv&  \left( 1- {2M\over r}\right)^{-1} (r^2 h_{AB} +r C_{AB} )= r^2h_{AB} + O(r).
\eea{m4a}

A  light ray moving on a closed loop returns to the origin after a time~\cite{psz}
 \beq
 \Delta u= \int d\lambda\left[ \sqrt{ ((P_AP_B +Q_{AB}) { d\Theta^A \over d\lambda}{ d\Theta^B \over d\lambda}} 
 + P_A  { d\Theta^A \over d\lambda} \right].
  \eeq{m5}
  
The time delay between two light rays moving along the same path in opposite directions is 
 \beq
 \Delta_+u -\Delta_- u= 2\int d\lambda  {d\Theta^A \over d\lambda} \left[ \hat{P}_A  + {2\over 3r} u(\lambda)
 \partial_A M \right] ,
 \eeq{m6}
 where we defined $\hat{P}_A\equiv P_A -(2/3r)u \partial_A M$.  

In a radiative region where the Bondi news tensor $N_{AB}$ is nonzero and $C_{AB}$ is generic,  eq.~\eqref{m6} 
is path-dependent, because all quantities appearing in it depend explicitly on $u$, hence on $\lambda$. Moreover,
the contribution to the delay due to $N_A$ is subdominant in $1/r$. 

In a nonradiative region, where $N_{AB}=0$, the mass aspect $M$, the angular momentum aspect $N_A$ and the shear $C_{AB}$ are $u$-independent; hence $\hat{P}_A$ is $u$-independent so it does not have any explicit 
dependence on $\lambda$. The affine parameter enters in all these quantities 
only implicitly, through $\Theta^A(\lambda)$.
Moreover it can be shown that $C_{AB}= -2(D_AD_B -\frac{1}{2} h_{AB} D^2)C$ --see e.g.~\cite{str13}.
 {The function $C$ parametrizes the shear in nonradiative regions and has been called by various names in the literature. We will call it --somewhat improperly-- the {\em boundary graviton}. Its $l=0,1$ harmonics are arbitrary since
 $C_{AB}$ does not depend on them, while the other ones are $u$-independent.}
 
Thanks to the form the shear tensor, $D^BC_{AB}= -D_A (D^2+2)C$ so the $O(r^0)$ term in $\hat{P}_A$ is a gradient
--and obviously $u$-independent. It is also convenient to make the choice $u=\lambda$.  
These properties  imply that the time delay in nonradiative regions for a closed path starting at $u=u_1$ and ending at $u=u_2$ is
 \bea
\Delta_+u -\Delta_- u &=& \frac 2 r \int d\Theta^A \wedge d\Theta^B D_A Q_B +  {4\over 3r} \int_{u_1}^{u_2} du u 
 {dM \over du },
\label{m7} \\
Q_A &\equiv&  \frac{2}{3}N_A   -M \partial_A (D^2+2)C .
\eea{m8}
Notice that $C$ and $M$ can be
 measured locally independently of $N_A$, so the time delay~\eqref{m7} can be used in principle to measure $N_A$.

In nonradiative regions supertranslations leave $M$ invariant, act on $N_A$ as in~\eqref{m1} and on $C$ as 
$C\rightarrow C' =C- f$. So the physical, observable time delay~\eqref{m7}
transforms nontrivially under a generic supertranslation.

{Another measure of  $N_A$ is through the well known Lense-Thirring (or frame-dragging) effect~\cite{lt1}. 
A thorough
explanation of the effect can be found for instance in~\cite{lt2}. The effect scales as $O(r^{-3})$ and requires to compute the
rate of precession of a gyroscope in a given reference frame --e.g. in the ``fixed stars'' frame. Unless the frame is carefully
chosen, the frame-dragging effect may be subdominant compared to other ones, which can appear at $O(r^{-2})$. 
On the other hand, those same effects can be used to measure the {\em change} in $N_A$ due to gravitational radiation
through spin memory~\cite{psz} or orientation memory~\cite{os1,os2}  effects. It can be shown that 
even in these cases a supertranslation affects  observable quantities, such as the angular velocity
 of a freely-falling gyroscope.
}

\section{Ambiguity in the definition of $N_A$} 
Supertranslations transform $N_A$ as in eq.~\eqref{m1}, so Lorentz charges transform for generic mass aspect. When the difference $\Delta M(\Theta)= M(\Theta,+\infty ) - M(\Theta,-\infty)$ is nonzero, the Lorentz charge fluxes also change.
The simplest way to obtain a supertranslation-independent definition of $\Delta J_Y$ and $J_Y|_(u=-\infty)$ is to select a 
convenient supertranslation frame. The shear transforms as 
\beq
C_{AB}(\Theta,u) \rightarrow C_{AB}'(\Theta,u')= C_{AB}(\Theta,u+f(\Theta) ) + 2D_AD_B f - h_{AB}D^2 f .
\eeq{m9}
The condition $D^AD^BC_{AB}=0$ can be used to select a supertranslation frame in the nonradiative region 
$u\rightarrow \pm \infty$ where we compute the flux  $\Delta J_Y$ and where $C_{AB}=-2 (D_AD_B  -\frac{1}{2}h_{AB}D^2) C$. There,
thanks to~\eqref{m9} we find
\beq
D^AD^B C_{AB} =-D^2(D^2+2)  C\rightarrow -D^2(D^2+2) C +D^2(D^2 +2)f ,
\eeq{m10}
so for any value of $C$ we can find a supertranslation frame where $D^A D^B C_{AB}=0$.
Expanding $C$ and $f$ in spherical harmonics as $C=\sum_{l=0}^\infty \sum_{m=-l}^l C_{lm} Y_{lm} $, 
$f=\sum_{l=0}^\infty \sum_{m=-l}^l f_{lm} Y_{lm} $ the operator $D^2(D^2+2)$ diagonalizes and the equation for the 
supertranslation frame becomes $l(l+2)(l^2-1) (C_{lm} - f_{lm})=0$. Notice that the $l=0,1$ harmonics of $f$ remain undetermined.

The aspect $N_A$, on the other hand, depends also on the $l=1,0$ harmonics of $f$. If we set them to zero we break Lorentz
covariance as it can be seen as follows.
Let us consider to be definite $N_A$ at $\mathscr{I}^+_-$.  A generic supertranslation frame with $D^AD^B C_{AB}=
\sum_{l=0}^\infty \sum_{m=-l}^l  l(l+2)(l^2-1) C_{lm} Y_{lm}\neq 0$ can be transformed into the ``reference frame'' 
$D^AD^BC_{AB}=0$ by setting $f_{lm}=C_{lm}$ for $l>1$. We could also set $f_{1m}=f_{00}=0$. An infinitesimal Lorentz transformation acts  on supertranslations as follows~\footnote{This formula is derived for instance in~\cite{stro-ir}.}
\beq
f(\Theta) \rightarrow f'(\Theta') = f(\Theta) -\delta_Y^{-1/2} f, \qquad
\delta^{w}_Y F \equiv w D\cdot Y F + Y\cdot D F .
\eeq{m11}
This formula says that $f$ transforms under $SL(2,C)$  as a primary of conformal weight $w=-1/2$. It can also be proven that
the mass aspect $M(\Theta)$ transforms  as a primary of conformal weight $w=3/2$. A boost can be written as $Y^A=D^A\psi$ with $\psi$ an $l=1$ harmonic obeying $D^2\psi =-2 \psi$. So, for an $l=2$ supertranslation,  obeying
 $D^2 f= -6 f$, we find
 \beq
 \delta_Y^{-1/2}f= -{1\over 2} f (D^2 \psi)  +{1\over 2} D^2 (f \psi ) + f\psi + 3 f\psi .
 \eeq{m12}
 The product $f\psi$ contains an $l=1$ harmonic on which $D^2(f\psi)|_{l=1}= -2 (f\psi)|_{l=1}$. Therefore 
 the boost of a supertranslation is an $l=1$ supertranslation, which is in fact just a nonvanishing  spacetime translation. 
 We can rewrite eq.~\eqref{m12} as an explicit statement on the BMS algebra, since when we  expand a supertranslation in spherical harmonics as $S=\sum_{lm} f_{lm}S_{lm}$ we can rewrite eq.~\eqref{m12} as the commutator of a 
 Lorentz boost $L_Y$ and a supertranslation,
 \beq
 [L_Y,S]=\sum_{lm}f_{lm}  [L_Y,S_{lm}]= \sum_{lm} \delta_Y^{-1/2} f|_{lm} S_{lm}.
 \eeq{m13}
 Here $...|_{lm}$ is the projection over a spherical harmonic, namely $g|_{lm}\equiv \int d^2\Theta \sqrt{h} g Y_{lm} $.
 
 So the choice $f_{1m}=f_{00}=0$ explicitly breaks invariance under Lorentz boosts because an $l>1$ supertranslation
 can be ``boosted down'' to an $l=1$ Poincar\'e algebra translation. We therefore need a covariant prescription to
 define the S- and P-wave harmonics of $f$, which appear explicitly in eqs.~\eqref{m1} and~\eqref{2}. This prescription was given
  in~\cite{3puzz}; here we will present it in a slightly more compact form and we will extend it to the case in which the final rest frame of a gravitationally radiating system at $\mathscr{I}^+_+$  is Lorentz-boosted w.r.t. the initial frame at $\mathscr{I}^+_-$.
 
 The key observation of~\cite{3puzz} was that {\em any} covariant two-index antisymmetric tensor, be it charge or flux, and which 
 is invariant under $l>2$ supertranslations must necessarily be invariant under the standard space-time translations of the 
 Poincar\'e group. The proof follows from the Jacobi identities of supertranslations. Schematically, if we call $S$ a supertranslation,
 $L$ a Lorentz transformation and $J_{\mu\nu}$ the covariant tensor, we have 
 \beq
 [[J_{\mu\nu},S],L]+ [[L,J_{\mu\nu}],S] +[[S,L],J_{\mu\nu}]=0 .
 \eeq{m14}
 The first term vanishes by assumption, the second is proportional to $J$ by covariance, and vanishes by our assumption of
 supertranslation invariance, hence the last term must vanish. On the other hand, we just proved in~(\ref{m12},\ref{m13}) 
 that $[S,L]$
 contains a generically nonzero translation. The upshot of this analysis is that a supertranslation-invariant and Lorentz-covariant tensor
 is also invariant under translations. Both Lorentz charges and their fluxes {\em can} change under translations.
 Angular momentum changes by the elementary formula $\vec{J} \rightarrow \vec{J} + \vec{a} \times \vec{P}$, 
 the boost charge by the
 formula $\vec{K} \rightarrow \vec{K} + E \vec{a}$ and their fluxes too change, if either the total momentum $\vec{P}$ or the total energy 
 $E$ of the radiating system change from $\mathscr{I}^+_-$ to $\mathscr{I}^+_+$. 
 
 So, a supertranslation-invariant angular
 momentum aspect is in reality a quantity defined with respect to a specific coordinate system. The challenge is to define such 
 system using only the asymptotic metric data~\eqref{1}. 
 
 \section{A covariant prescription for the (super) translation frame}
 
 The center of mass (CM) of a nonrelativistic system of particles with masses $m_a$, $a=1,...,N$  is $\vec{x}_{CM}= \sum_{a=1}^N M_{CM}^{-1} m_a \vec{x}_a $ with  $M_{CM}\equiv \sum_{a=1}^N m_a$. 
 In rest frames, where the total momentum 
 $\vec{P}_T\equiv \sum_{a=1}^N \vec{p}_a=0 $,  $\vec{x}_{CM}$ is
 time independent. The CM frame is by definition the rest frame  for which $\vec{x}_{CM}=0$. 
 For relativistic particles and and fields, rest frames are defined by $\vec{P}_T=0$ and by the 
 vanishing of boost Lorentz charges  $J_{0i}=0$, $i=1,2,3$. The covariant prescription for the CM frame is thus
 \beq
 J_{\mu\nu}P^\nu=0.
 \eeq{m15}
 In this form, the prescription requires to know only the Lorentz charge and the total 4-momentum of the system, $P^\mu$.
 This prescription can be translated into formulas involving only asymptotic metric quantities by noticing that the Lorentz charges
 defined in~\eqref{2} map $V$, the space of  Conformal Killing Vectors (CKVs), into real numbers. So the Lorentz charges belong to $V^*$, the dual vector space of $V$. This means that
 the duals of a Lorentz charge is an element of $V$, so it is a CKV.
 Concretely, we choose a basis  for $V$, say $Y_\alpha$, $\alpha=1,..6$ and write the CKV 
 \beq
 Y^A(J) = \sum_\alpha Y^A_\alpha J_{Y_\alpha}(-\infty) .
 \eeq{m16}
 We are resolving the ambiguity in the initial $N_A$ first, hence in this formula we used the initial angular momentum.
 
 In terms of $Y^A(J)$ eq.~\eqref{m15} is
 \beq
 \delta_{Y(J)}^{3/2}M|_{lm} =0, \quad \mbox{for } l=0,1.
 \eeq{m17}
 In the rest frame where $M|_{l=1}$ vanishes this equation reduces to 
 \beq
 \int d^2\Theta \sqrt{h} Y^AN_A = 0, \quad Y^A=\mbox{ pure boost}=D^A \psi, \quad D^2\psi=-2 \psi.
 \eeq{m17a}
 This is indeed $J_{i0}=0$ in vector notations.
 To see how this equation fixes the $l<2$ harmonic components of $f(\Theta)$ we first use the $l>1$ harmonic components of $f(\Theta)$ to reach the supertranslation frame where $D^A D^B C_{AB}=0$. Then we perform a translation on $N_A$, 
 $f(\Theta)=f_{00} + \sum_{m=-1}^1 f_{1m} Y_{1m}(\Theta)$, and combine eq.~\eqref{m17a} with 
 eq.~\eqref{m1}  to obtain, after some integrations by part,
 \beq
0=\int d^2\Theta \sqrt{h} D^A \psi (N_A - 3MD_Af - D_AM f)=\int d^2\Theta \sqrt{h} [D^A \psi N_A  + M\psi D^2f  - 4 M\psi  f -MD^2(\psi f)] .
\eeq{m20}
The scalar harmonic of $f(\Theta)$ does not appear here since $D^2 f_{00}=0$ and 
$\int d^2\Theta \sqrt{h} \psi M f_{00}\propto M|_{l=1}=0$ in the rest frame. Choosing $\psi=Y_{1m}$ and integrating by parts the last term in~\eqref{m20}, we note that the $l=2$
harmonics of the mass aspect satisfy $D^2M_{l=2}=-6M|_{l=2}$ so their contribution to~\eqref{m20} cancels. The $l=1$ harmonics are then 
\beq
f_{1-m} =-{1\over 6 M_{00}}\int d^2\Theta \sqrt{h}  Y_{1m} D^AN_A .
\eeq{m20a}
Notice that $f_{1-m}$ depends only on  the gradient part of $N_A$, as it should, because in the rest frame  
boost charges change under shift of the center of mass coordinates while angular momentum doesn't.

 As we noticed earlier, $M(\Theta)$ transforms as an $SL(2,C)$ primary of conformal weight $3/2$ and this means that 
the constraint given in~\eqref{m17} depends only on the $l=0,1$ harmonics of $M$, which are the 4-momentum $P^\mu$. We 
have seen this explicitly in the CM frame in \eqref{m20a}. To
check this property in a generic Lorentz frame we write the transformation of $M|_l$, the $l$-harmonic of $M$. We write $Y^A(J)=\epsilon^{AB} D_B\phi + D^A \psi$. 
For a rotation $\epsilon^{AB}D_B \phi$, the property is obvious; for a boost $D^A \psi$
  \beq
  \delta^{3/2}_{D\psi} M|_l  = {3\over 2} D^2\psi  M|_l  + D\psi \cdot D  M|_l = -3\psi  M|_l + D\psi \cdot D M|_l  = {1\over 2} \left[ D^2 -4 +l(l+1)\right](\psi M|_l)  .
  \eeq{m18}
The product $\psi M_l$ can appear in eq.~\eqref{m17} only when it contains scalar or vector harmonics. Since $\psi$ is an $l=1$ 
harmonic, the only $M|_l$ with $l>1$ that can appear in~\eqref{m17} is $M|_{l=2}$, but for $l=2$ we find
 $\delta^{3/2}_Y M|_{l=2}={1\over 2} \left[ D^2 +2 \right](\psi  M|_{l=2})  |_{l=1}=0$.
 The property that the formula in~\eqref{m17} depends only on the first two harmonics of $M$ 
 follows directly from the structure of the Lorentz algebra. Under Lorentz transformations the 
4-momentum $P^\mu$ transforms as a 4-vector. The $l\leq 1$ harmonics in the mass aspect are linear combinations of $P^\mu$. 
By expanding in spherical harmonics we find that the general structure of a Lorentz transformation of the mass aspect is
\beq
\delta^{3/2}_Y M = 
\sum_{l=0}^\infty \sum_{-l\leq m \leq l} \sum_{l'=0}^\infty \sum_{-l'\leq m' \leq l'} M_{l'm'} {R_Y}_{lm}^{\;\; l'm'} Y^{lm}(\Theta) .
\eeq{m18a}
Since $P^\mu$ transforms into linear combinations of $P^\mu$ only, the coefficients ${R_Y}_{lm}^{\;\; l'm'}$ that transform higher
harmonics $M_{l'm'}$ into $l=0,1$ ones must vanish: ${R_Y}_{lm}^{\;\; l'm'}=0$ for $l'>1$, $l\leq 1$. 
We can denote with the shorthand $\underline{M}$ the vector of coefficients
$M_{lm}$ in its expansion in spherical hamonics. Then the Lorentz  transformation law in matrix form and with obvious notations is
\beq
\delta^{3/2}_Y \underline{M} = \begin{pmatrix} R_{l>1, l'>1} & R_{l>1,l'\leq 1} \\ 0 & R_{\leq 1, l'\leq 1} \end{pmatrix} \underline{M}.
\eeq{m18b}
 
Eq.~\eqref{m17}  is a manifestly covariant prescription that fixes the ambiguity in the choice of the $l=0,1$ harmonics of  $f(\Theta)$ in~\eqref{m2}. It is compatible with the frame choice $D^AD^B C_{AB}=0$ because a Lorentz transformation of $f|_l$, $l=0,1$ 
does not produce $l>1$ terms. So it does not change the harmonics which are uniquely determined by 
$D^AD^B C_{AB}=0$. This property is obvious for rotations and it is seen for a boost $Y^A= D^A \psi$ using
 \beq
 \delta_Y^{-1/2}f= -{1\over 2} f (D^2 \psi)  +{1\over 2} D^2 (f \psi ) + f\psi + {l(l+1)\over 2} f\psi =
 {1 \over 2} [D^2 + 4 +l(l+1)](f\psi). 
 \eeq{m19}
The only possible higher harmonic in~$\delta_Y^{-1/2}f$ which is produced by the $l=0,1$ harmonic components of $f$ is an  $l=2$. It arises from the product $f|_{l=1} \psi$, but eq.~\eqref{m19}  shows that such a term vanishes.

\section{A covariant prescription for the angular momentum and mass dipole fluxes}

The prescription given by~\eqref{m17} does not fixes the $l=0$ harmonic of $f$, but this is not the end of the story. We have worked
so far only with quantities defined at $u=-\infty$. We need to make a choice of (super)translation frame also for the final 
$N_a(+\infty)$. Various possibilities exist. The importance of such choices for the definition of angular momentum and mass
dipole fluxes has been especially emphasized in~\cite{vv,rvw,mww,v25}.
\begin{enumerate}
\item We could define $\Delta N_A$ using for both $N_A(\pm \infty)$ the frame choice $D^A D^B C_{AB}(-\infty)=0$ and 
$\delta_{Y(J)}^{3/2}M|_{l=0,1} (-\infty)=0$. The final frame generically has $D^A D^B C_{AB}(+\infty)\neq 0$ because of memory
effects (a.k.a. soft gravitational hair implant~\cite{hps2}). So generically an initial round metric $C_{AB}=0$ is deformed to
$C_{AB}=-2(D_AD_B -\frac{1}{2}h_{AB} D^2)C\neq 0$. With this choice the angular momentum flux of radiating sources begins at 
$O(G^3)$. This choice is the choice made in~\cite{comp,jkp}.
\item We could define 
$\Delta N_A$ using the frame choice $D^A D^B C_{AB}(-\infty)=0$ and 
$\delta_{Y(J)}^{3/2}M|_{l=0,1} (-\infty)=0$ for $N_A(-\infty)$ but instead use $D^A D^B C_{AB}(+\infty)=0$ for $N_A(+\infty)$. This is the choice made in~\cite{yau21a,yau21b} and in most literature computing gravitational radiation from physical 
sources~\cite{dam,bd,conf1,conf2,conf3,conf4,conf5,man,heis}. Here, all memory effects are discarded and $N_A(+\infty)$  is computed using the round metric with $C_{AB}(+\infty)=0$. The angular momentum flux begins at $O(G^2)$.
\end{enumerate}
Either with choice 1 or 2 we need to find a new covariant prescription to fix the translation frame. This is clear by considering the 
initial rest frame, where the $l=1$ harmonic of $M(\Theta,-\infty)$ vanishes and $f_{00}$ is undetermined.

On the other hand, in the presence of gravitational radiation, generically we have $M|_{l=1}(+\infty)\neq 0$. The
choice $\delta^{-1/2}_Y(J)M_{lm}(-\infty)=0$ then leaves the freedom to shift $f_{00}$. Under 
$f(\Theta)\rightarrow f(\Theta) + constant$ the final mass dipole $J_{D\psi}(+\infty)=\int d^2\Theta \sqrt{h} D^A \psi N_A (+\infty)$ 
changes as 
\beq 
\int d^2\Theta \sqrt{h} D^A \psi N_A (+\infty) \rightarrow \int d^2\Theta \sqrt{h} D^A \psi N_A (+\infty) +
constant  \int d^2\Theta \sqrt{h} \psi M .
\eeq{m21}
Setting $\psi$ equal to the $l=1$ spherical harmonic $Y_{1m}$ we get $J_{DY_{1m}} (+\infty) \rightarrow J_{DY_{1m}} (+\infty) + constant M_{1-m}(+\infty)$. We can fix this freedom by imposing for instance $J_{DM|_{l=1}}(+\infty)=0$. 

In tensor notation the meaning of the freedom to shift $f_{00}$ becomes clearer: it means that the final mass dipole $J^f_{0i}$ is
ambiguous: when the initial system in in the rest frame, the final mass dipole can be shifted by an amount proportional to the final linear momentum $P^{f\, i} $. This is the ambiguity that follows from the classical mechanics definition 
$J_{0i}= Ex^I_{CM} - P^i t_{CM}$, where it corresponds to a shift in the origin of time $t_{CM}$. This ambiguity can be fixed by choosing $J^f_{0i}P^{f\, i}=0$. In covariant notations, if we call $P^i_\mu$ the initial 
total 4-momentum and $P^f_\mu$ the final 4-momentum we can write this condition in manifestly covariant form  as
\beq
P^{i\,\mu} P^{f\, \nu} J^f_{\mu\nu}=0 .
\eeq{m22}
Intriguingly, in the initial rest frame it is precisely the component of the mass aspect proportional to $\vec{P}^f$ that suffers from an infrared divergence
at $O(G^4)$, as shown in~\cite{hr}.

\section{Difference between two consistent definitions of Lorentz charges}\label{diff}
By construction {\em any} definition of 
supertranslation-invariant Lorentz generators, $J^{inv}_Y$,
makes them commute with the $l>1$ harmonics of the mass aspect at $\mathscr{I}^+_-$, $M(\Theta, u)|_{u=-\infty}\equiv M^-(\Theta)$,
 \beq
 [J^{inv}_Y,M^-(\Theta)|_{l>1}]=0. 
 \eeq{madd1}
 On the other hand, the transformation law of $M^-$ can be found by
 performing a Lorentz transformation on the asymptotic 
 metric~\eqref{m1}. It is given by eq.~\eqref{m11}  with $w=3/2$, from which it is
 obvious that neither boosts nor rotations vanish on $M_{l>1}$.
 
 We must then conclude that
  while the supertranslation invariant Lorentz generators are useful quantum operators -- in fact essential for defining unambiguously
   the angular momentum of the vacuum as well as other quantum numbers-- they should not be used to generate Lorentz 
   transformations on the fields.
   In fact, the supertranslation-invariant Lorentz charges
   are elements of the
   {\em universal enveloping algebra} of an enlarged BMS algebra that includes logarithmic
   supertranslations. Their explicit form is given e.g. in eq.~(9.7) of
   ref.~\cite{henn1} and eqs.~(18,19,20,23) of~\cite{henn2}, eq.~(6) of~\cite{yau21a}, eq.~(1.8) of ~\cite{yau21b}, eq.~(5.13) of~\cite{comp} eq.~(3.23) of~\cite{jkp}. They generate a Lorentz transformation
   specified by the conformal Killing vector $Y$ {\em
     plus a $Y$-dependent transformation} generated by supertranslations and by the boundary graviton $C^-$, which is canonically conjugated to $M^-$. 

\section{A concrete example: the metric of $n$ point particles}
The general analysis done in the previous sections can be illustrated nicely by the
 example of  the linearized metric sourced by the stress tensor of $n$ point particles, computed at order  $O(G)$.
The explicit form of the metric in Cartesian coordinates $ x^{\mu} = (x^0, x^i)$ is 
 \begin{gather}\label{ppmetric}
 	g^{\mu\nu} = \eta^{\mu\nu} - 4G \sum \frac{m}{\Gamma(x)} \left(v^{\mu}v^{\nu} + \frac{1}{2}\eta^{\mu\nu}\right),  \\
 	\Gamma(x) = \sqrt{(x^{\mu} - c^{\mu})(x^{\nu} - c^{\nu})\Pi_{\mu\nu}} \, ,  \\
 	\Pi_{\mu\nu} = \eta_{\mu\nu} + v_{\mu}v_{\nu} \, , \quad v^{\mu}\Pi_{\mu\nu} = 0 \, . \
 \end{gather}
 In this metric the $\sum$ sign is a sum over all the particles $i$ each having their own mass $m_i$ and momentum $p^\mu_i=m_i v^\mu_i$; however, unless explicitly shown, the index $i$ is suppressed everywhere to avoid clutter but nonetheless it is assumed wherever the $\sum$ sign appears. Since we are only considering $O(G)$ effects, all index operations are done via Minkowski metric $\eta_{\mu\nu} = \operatorname{diag}(-1, 1, 1, 1)$, particles are moving on straight lines with the constant velocity $v^\mu_i$ and their position at the proper time $\tau$ is given by
 \begin{gather}
 	x_i^{\mu}(s) = c_i^{\mu} + v_i^{\mu} \tau \, , \quad v_i^2 = -1, 
 \end{gather}
where $c^{\nu}$ is a the position of the $i$-th particle at $\tau=0$. 

\subsection{Angular momentum aspect}
In this subsection we find the total angular momentum and the angular momentum aspect $N_A$ for this metric by finding a diffeomorphism that  puts the metric (\ref{ppmetric}) into the Bondi-Sachs gauge. In this gauge the metric has the form (\ref{1}) and the angular momentum aspect $N_A$ is most easily read from the coefficient of $O(1/r)$ term of the $g_{uA}$ of the metric.
\begin{align}
	g_{uA}= \frac{1}{2}D^AC_{AB}+ \frac{1}{r}\left(
	\frac{2}{3}\left(N_A+u \partial_A M \right) - \frac{1}{16} \partial_A \left(C_{BD}C^{BD}\right)\right)
\end{align}
The explicit expressions for the mass aspect, angular momentum aspect and the shear tensor has been worked out in  Appendix A. The following expressions are linear in $G$, this means that  multiparticle expressions are simply given
by  the sum over the contribution of each single particle. The single-particle contribution can be conveniently 
calculated by aligning the particle velocity with the $z$ axis.
\begin{align}
	&C_{\theta\theta}=\frac{2 G m  v^2 \sin ^2(\theta )}{\sqrt{1-v^2} (1-v \cos (\theta ))},\\
	&C_{\phi\phi}=-C_{\theta\theta} \sin^2(\theta),\\
	&C_{\theta\phi}=C_{\phi\theta}=0,\\
	&M=\frac{G m}{(1-v^2)^{\frac{3}{2}} (1-v \cos (\theta ))^3},\\
	&N_\theta=-\frac{3G m \left(1-v^2\right)^{3/2}}{(v \cos (\theta )-1)^4} \bigg(\sin (\theta ) (z_0- vt_0)+\big(x_0 \cos (\phi )+y_0 \sin (\phi )\big) (v-\cos (\theta ))\bigg),\\
	&N_\phi=-\frac{3 G m \left(1-v^2\right)^{3/2}}{(v \cos (\theta )-1)^3} \sin (\theta ) \Big(y_0 \cos (\phi )-x_0\sin (\phi )\Big).
\end{align}
Now a simple application of \eqref{2} gives us the mechanical angular momentum of the system.
\begin{align}
	\vec{J}=(J_x,J_y,J_z)=\frac{m v}{\sqrt{1-v^2}}(y_0,-x_0,0) .
\end{align}
This simple answer was certainly expected although the way we arrived at this result was complicated, but for a more complex gravitational system the derivation described here must be used.

Under a supertranslation $f$, the shear tensor transforms as 
\begin{equation}
\delta_fC_{AB}=-2(D_AD_B-\frac{1}{2} h_{AB}D^2)f.
\end{equation}
We can set the shear tensor to zero by application of a supertranslation. It is easy to check that the supertranslation $f$ that we are looking for is
\begin{equation}
	f(\theta,\phi)=-2Gm(n\cdot v) \log(-n\cdot v), \quad n=(1,\hat x), \quad \hat x = r^{-1} \vec{x} .
\end{equation}
It should be noted again that this term gives a correction of order $G^2$ to the angular momentum aspect and therefore it can be discarded here. The general expression for the multiparticle metric is a sum over the contribution of each of the particles, plus an extra term coming from the contribution of the center of the mass, 
which is necessary from the covariance of the shear tensor. This expression is worked out in the next subsection.   

 \subsection{Covariance of the mass and angular momentum aspect}
 We consider two frames that are related by a boost $\vec{v}$.
The null vectors $n_1^\mu$ and $n_2^\mu$ are defined as 
\begin{align}
	n_1^\mu=(1,\sin(\theta_1)\cos(\phi_1),\sin(\theta_1)\sin(\phi_1),\cos(\theta_1)),\\
	n_2^\mu=(1,\sin(\theta_2)\cos(\phi_2),\sin(\theta_2)\sin(\phi_2),\cos(\theta_2)).
\end{align}  
Without loss of generality we can assume that the $z$-axis in frames $1$ and $2$ are chosen parallel to $\vec{v}$, since
covariance under rotations is easy to check. In the first frame, let's consider a particle moving with velocity $u_1^\mu$, then $u_1 \cdot n_1$ is Lorentz invariant. Now if $u_2$ is the velocity of the particle in the boosted frame then we have,
\begin{align}\label{n.v}
	n_1\cdot u_1=-(n_1\cdot v)(n_2\cdot u_2).
\end{align}
By construction $u_2$ is the relativistic sum of the two velocities
$u_1$ and $v$, hence we found a relation between the scalars $n\cdot v$ in different frames.
With the help of this relation we find that the mass aspect for $n$ 
particles moving with velocities $v_1,...,v_n$ transforms as
\begin{align}\label{masstransformation}
	M_1(\theta_1,\phi_1)=\sum_{i=1}^{n}-\frac{G m_i}{(n_1\cdot v_i)^3}=-\frac{M_2(\theta_2,\phi_2)}{(n_1\cdot v)^3}.
\end{align}
This transformation rule is the manifestation of the fact that the mass aspect is a scalar of weight $\frac{3}{2}$.
The mass aspect carries all information about the mass multipoles of the system, in particular the $l=0$ harmonic gives the total energy and the $l=1$ ones give the total momentum of the system. The $l=0,1$ harmonics can be combined to construct the four momentum $P^\mu$ of the system. In the center of mass frame $P^\mu=(m,\vec{0})$ and in the boosted frame we expect it to be $P^\mu=\frac{m}{\sqrt{1-v^2}}(1,\vec{v})$. Now it is easy to check from the relation (\ref{masstransformation}) that if we set the mass aspect $M_2(\theta_2,\phi_2)$ equal to its value in the rest mass $G m$, then we have
\begin{align}\label{masstransformation2}
	M_1(\theta_1,\phi_1)=\frac{G m}{\sqrt{1-v^2}}+\frac{Gm}{\sqrt{1-v^2}}\sum_{m=-1}^{1} v^m Y_{1m}(\theta_1,\phi_1)+\sum_{l>1,m=-l}^{m=l}M_1^{lm}Y_{lm}(\theta_1,\phi_1) .
\end{align}
This indeed agrees with our earlier observation that the two first harmonic of the mass aspect represent the four-momentum of the system and do not get mixed with  higher multipoles.

Now let us study the ``boundary graviton'' $C$, that we introduced after eq.~\eqref{m6} and is defined as 
\begin{align}\label{rj1}
	C_1(\theta_1,\phi_1)&=\sum_{i=1}^{n}\log(-n_1\cdot v_i) G m_i (-n_1\cdot v_i)+\log(-n_1\cdot v_{CM})\sum_{i=1}^{n}G m_i (n_1\cdot v_i).
\end{align}
Our supertranslation frame choice is  $C+f=0$, with the proviso  that the redundant $l=0,1$ components of either $f$ or $C$ are fixed by~\eqref{m15}. The transformation rule for the boundary graviton can now be worked out similarly to those of the
mass aspect and we have 
\begin{align}
	C_1(\theta_1,\phi_1)&=-(n_1\cdot v)C_2(\theta_2,\phi_2),
\end{align}
which could have been easily obtained also by recalling that the boundary graviton is a scalar of weight $-\frac{1}{2}$.

\subsection{First two modes of boundary graviton in any frame}
Our construction highlights the importance of the $l=0$ and $l=1$ modes of the boundary graviton. The procedure begins by identifying the zero mode in a given frame, followed by a similar procedure to determine the $l=1$ mode. Rotations, corresponding to a conformal factor equal to unity, leave the boundary graviton unaffected. In contrast, boosts generate an angle-dependent conformal factor, inducing a nontrivial mixing between modes of different $l$. To simplify the calculations, we first introduce the parameter $\beta \equiv \frac{1}{v}$ and assume that the boost is along the $z$-axis in both the first and second frames, therefore for $l\neq 0$ harmonics only their $m=0$ components can receive nonzero contribution. Rather than presenting the full derivation, we directly state the final expression for the \( l = 0 \) mode of the boundary graviton in frame 2, expressed in terms of the modes in frame 1.
\begin{align}
	C_2\vert_{l=0}&=\frac{1}{2v^3\gamma_v^3}\left[\sum_{l=0}^{\infty}C_1\vert_{l0}\frac{\partial}{\partial \beta^2}Q_l(\beta)\right].
\end{align} 
Where $Q_l$ is the Legendre function of second kind. This explicitly shows that even when the boundary graviton $C_1$ lacks \textbf{S}- and \textbf{P}-waves in frame~$1$, they can still be generated in frame~$2$ from higher multipole modes ($l \geq 2$).

Similarly for the $l=1$ mode we have
\begin{align}
	C_2\vert_{l=1,m=0}&=-\frac{3}{2v^3\gamma_v^3}\sum_{l=0}^{\infty}
	C_1\vert_{l0}\left[\beta\frac{\partial^2}{\partial \beta^2}Q_l(\beta)+\frac{\beta^2-1}{3}\frac{\partial^3}{\partial \beta^3}Q_l(\beta)\right].
\end{align} 
By using the above technique one can find exact expressions also  for $C_2\vert_{l=1,m=\pm1}$. For instance, for $m=+1$
we find
\begin{align}
	C_2\vert_{l=1,m=1}
	&=-\frac{1}{4\gamma_v^4 v^4}\frac{\partial^3}{\partial \beta^3}\sum_{l=0}^{\infty}C_1\vert_{l1}\left[(1-\beta^2)\frac{\partial}{\partial\beta}Q_l(\beta)\right].
\end{align} 

\subsubsection{Large $\beta$ limit}
It is interesting to derive an expression in the large $\beta$, or equivalently small $v$ limit, corresponding to slowly moving frames. In this limit, the behavior of the Legendre function is approximatec by
\begin{equation}
	Q_l(\beta)=\frac{\sqrt{\pi}\Gamma(l+1)}{\Gamma(l+\frac{3}{2})(2\beta)^{l+1}}=\frac{2^{l+1}l!(l+1)!}{(2l+2)!\beta^{l+1}}.
\end{equation}    
Therefore in this limit we have,
\begin{align}
	C_2\vert_{00}&=\sum_{l}\frac{2^l(l+1)!(l+2)!}{(2l+2)!}v^lC_1\vert_{l0}=C_1\vert_{00}+v C_1\vert_{10}+O(v^2) ,\\
	C_2\vert_{10}&=\sum_{l}\frac{2^l(l+1)!(l+2)!l}{(2l+2)!}v^{l-1}C_1\vert_{l0}=C_1\vert_{10}+2vC_1\vert_{20}+O(v^2),\\
	C_2\vert_{11}&=\sum_{l}\frac{2^{l-1}(l+1)!(l+2)!l(l+1)}{(2l+2)!}v^{l-1}C_1\vert_{l1}=C_1\vert_{11}+3vC_1\vert_{21}+O(v^2).
\end{align} 
This explicitly shows the mixing between modes with different $l$ caused by a boost.

\subsection*{Acknowledgments}
We would like to thank Marc Henneaux for useful discussions and comments. 
MP is supported in part by NSF grant PHY-2210349. 
MP would like to thank the Galileo Galilei Institute for Theoretical Physics for its kind hospitality and partial support through
 the INFN and the Simons Foundation during the early stages of this work.

\appendix
\section{Point particle metric in the Bondi gauge}
The Bondi-Sachs gauge is defined by
\begin{align}
	g_{rr}=g_{rA}=0,~~~ \partial_r \det(\frac{g_{AB}}{r^2})=0.
\end{align}
These conditions translate into the following equations for the new coordinates $(u,r,\Theta^A)$ in terms of the old coordinates $(u_M,r_M,\Theta^A_M)$ of Minkowski spacetime.
\begin{align}
	&(\nabla_\mu u)(\nabla^\mu u)=(\nabla_\mu u)(\nabla^\mu \Theta^A)=0,	\\
	&\det(\nabla^\mu \Theta^A,\nabla_\mu \Theta^B)=\frac{1}{r^4} \det(\Omega_{AB})^{-1}.
\end{align}
We can find a perturbative  solution of these equations to linear order in $G$. The zeroth order solution is trivially  $(u_M,r_M,\Theta^A_M)$. At linear order in $G$ we have 
\begin{align}
	&u=u_M+\delta u,\\
	&r=r_M+\delta r,\\
	&\Theta^A=\Theta^A_M+\delta \Theta^A.
\end{align}
By inserting these formulas in the Bondi-Sachs gauge condition, we obtain the equations for the shifts $\delta u$, $\delta r$ and $\delta \Theta^A$ \cite{vv}.
\begin{align}\label{diffeq}
	\partial_{r} \delta u &= -\sum \frac{2Gm}{\Gamma} (n \cdot v)^2  , \\
	\partial_{r} \delta\Theta^A &= \frac{1}{r^2} \Omega^{AB} \partial_B \delta u + \frac{1}{r} \sum \frac{4Gm}{\Gamma} (n \cdot v) \Omega^{AB} \partial_B (n \cdot v)  , \label{diffeq2}\\
	- 2  \delta r &= r D_A \delta \Theta^A - r \sum \frac{2Gm}{\Gamma} (\Omega^{AB} \partial_A (n \cdot v) \partial_B (n \cdot v) + 1) .\label{diffeq3}
\end{align}
Here we have introduced for clarity a null vector $n^\mu=(1,n^i)=\eta^{\mu\nu} \partial_\nu u$. As we will see this vector help us simplifying and representing the results.  
The factor $n\cdot v$ can be written most clearly and without loss of generality explicitly in terms of the coordinates if we rotate the axis so that the velocity is along the $z-$axis.  
\begin{equation}\label{n.v1}
	n\cdot v=n_\mu v_\mu=-\gamma(1-\vec{v}\cdot \vec{n})=-\frac{1}{\sqrt{1-v^2}}(1-v \cos(\theta)).
\end{equation}
Depending on the calculation at hand we will use either of the above forms for $n\cdot v$. For large radius we can find a perturbative solution to (\ref{diffeq}),(\ref{diffeq2}) and (\ref{diffeq3}) to the desired order in $\frac{1}{r}$. The solution can be represented most conveniently if we solve it first for a given order. To see this, one must start by expanding the $\Gamma$ function in the metric,
\begin{align}\label{gamma}
	\frac{1}{\Gamma}=\sum_{p=1}^{\infty}\frac{\Gamma^{(p)}}{r^p}.
\end{align}
 Since we only keep terms to linear order in $G$, the coordinate transformation and other important quantities like the mass
  aspect and angular momentum aspect, will also have this property. This means that if we restrict ourselves to the
  single-particle case, we can easily generalize it to the multi-particle case by summing over the contributions of each 
  particle. It is useful to write the explicit form of  $\Gamma^{(p)}$ for the first few $p$ for a single particle moving in the 
  $z-$direction that starts at $(x_0,y_0,z_0)$ at $t=t_0$.
\begin{align}\label{gamma1}
	&\Gamma^{(1)}=-\frac{1}{n\cdot v},\\
	&\Gamma^{(2)}=-\frac{1}{(n\cdot v)^3}\bigg(x_0 \cos(\phi)\sin(\theta)+y_0 \sin(\phi)\sin(\theta)+(1+\gamma n\cdot v)(u+\frac{z_0-vt_0}{v})\bigg),\\
	&\Gamma^{(3)}=g(\theta,\phi)+\frac{u^2 v^2 \sqrt{1-v^2}}{4 (v \cos (\theta )-1)^5} \Big(\left(v^2-3\right) \cos (2 \theta )-5 v^2+8 v \cos (\theta )-1\Big)\\
&\quad\quad~+\frac{u v \sqrt{1-v^2} }{2 (v \cos (\theta )-1)^5}\bigg(-\Big((t_0 v-z_0) \left(\cos (2 \theta )\left(v^2-3\right) -5 v^2+8 v \cos (\theta )-1\right)\Big)\rightarrow\nonumber\\
 &\quad\quad\quad\quad\quad\quad\quad\quad\quad\quad\quad\rightarrow+6 \left(v^2-1\right)  \sin (\theta )  (\cos (\theta )-v)\Big(x_0\cos (\phi )+y_0 \sin (\phi ) \Big)\bigg).
\end{align}
Here $g(\theta,\phi)$ is a function of the angles and initial position of the particle but is independent of the retarded time $u$. 
One can check by a direct calculation that this term does not contribute to either the angular momentum or mass aspect, therefore we do not write it explicitly.

Now we can find the solution to equations (\ref{diffeq}),(\ref{diffeq2}) and (\ref{diffeq3}) at each order in $\frac{1}{r}$ by 
using (\ref{gamma}). The solution to these equation to first order has been worked out in \cite{vv} and later expanded 
partially to second order in \cite{jp}. In \cite{jp} the aim was to find the mass aspect and define a supertranslation invariant angular momentum. 

For the case $p>1$ we can find the solution by an integration over $r$,
\begin{align}
	&\delta u= \frac{2Gm (n\cdot v)^2\Gamma^{(p)}}{(p-1)r^{p-1}},\\
	&\delta\Theta^a=-\frac{2Gm}{r^p}\Big(\frac{1}{p(p-1)}D^a(\Gamma^{(p)}(n\cdot v)^2)+\frac{(n\cdot v)^2}{p}D^a\Gamma^{(p)}\Big),\\
	&\delta r=\frac{Gm}{r^{p-1}}\Bigg(\frac{1}{p(p-1)}D^2\big(\Gamma^{(p)}(n\cdot v)^2\big)+\frac{1}{p}D_a\big((n\cdot v)^2D^a\Gamma^{(p)}\big)+\Gamma^{(p)}(1+D_a(n\cdot v)D^a(n\cdot v))\Bigg).
\end{align}
For $p=1$ the solution is quite simple,
\begin{align}
	&\delta u= -2Gm (n\cdot v)^2\Gamma^{(1)}\log(r),\\
	&\delta\Theta^a=2GmD^a(n\cdot v)(\frac{1-\log(r)}{r}),\\
	&\delta r=Gm(4\gamma+3 n\cdot v)-2m\log(r)(\gamma+n\cdot v).
\end{align}
It should be mentioned that here we set the constants of integration to zero, we refer to \cite{vv} for a discussion about the relevance of these constants. In short, the result is that some of the constants can be absorbed into other parameters or set to zero, but there remains also a part that corresponds to a supertranslation. Here we are not considering any extra supertranslation on top of the coordinate transformation needed to put the metric into the Bondi gauge, hence we can set these constants to zero. 
 
After coordinate transformation, the metric satisfies the Bondi gauge conditions and takes the form~(\ref{1}); therefore,
the components of the shear tensor $C_{ab}$, the angular momentum aspect $N_A$, and the mass aspect $M$ can  
all be found easily from the metric,
\begin{align}
	&C_{\theta\theta}=\frac{2 G m  v^2 \sin ^2(\theta )}{\sqrt{1-v^2} (1-v \cos (\theta ))},\\
	&C_{\phi\phi}=-C_{\theta\theta} \sin^2(\theta),\\
	&C_{\theta\phi}=C_{\phi\theta}=0,\\
	&M=-\frac{G m}{(n\cdot v)^3},\\
	&N_\theta=-\frac{3G m \left(1-v^2\right)^{3/2}}{(v \cos (\theta )-1)^4} \bigg(\sin (\theta ) (z_0- vt_0)+\big(x_0 \cos (\phi )+y_0 \sin (\phi )\big) (v-\cos (\theta ))\bigg),\\
	&N_\phi=-\frac{3 G m \left(1-v^2\right)^{3/2}}{(v \cos (\theta )-1)^3} \sin (\theta ) \Big(y_0 \cos (\phi )-x_0\sin (\phi )\Big).
\end{align}
Note that the quadratic term $D_a(C_{bc}C^{bc})$, which would normally be present in the angular momentum aspect, is not relevant here because it is $O(G^2)$.   


\end{document}